\documentclass[aps,prd,twocolumn,superscriptaddress,nofootinbib,amsmath,amssymb,floatfix,10pt]{revtex4-2}

\usepackage{graphicx}
\usepackage{booktabs}
\usepackage{tabularx}
\usepackage{xcolor}
\usepackage{xurl}
\usepackage{hyperref}
\graphicspath{{img/}}

\hypersetup{
  hidelinks,
  pdftitle={Stress-Testing DANTE under Detector Domain Shift: a Representation-Coherent Reanalysis of LIGO O4a},
  pdfauthor={Luca Cirfeta},
  pdfsubject={Sixth-version validation and correction of the DANTE O4a analysis},
  pdfkeywords={LIGO, detector characterization, transient noise, domain shift, unsupervised machine learning}
}

\newcommand{\robust}{\textsc{Robust}}
\newcommand{\ambiguous}{\textsc{Ambiguous}}
\newcommand{\background}{\textsc{Background}}
\newcommand{\vcell}[2]{\parbox[t]{#1}{\raggedright #2}}

\begin{document}

\title{Stress-Testing DANTE under Detector Domain Shift:\\
a Representation-Coherent Reanalysis of LIGO O4a}

\author{Luca Cirfeta}
\thanks{ORCID: \href{https://orcid.org/0009-0000-1235-3186}{0009-0000-1235-3186}}
\affiliation{Independent Researcher, Rome, Italy}
\date{\today}

\begin{abstract}
This representation-coherent Domain-Adaptive Network for Transient Evaluation
(DANTE) preprint stress-tests the inference made by an
unsupervised transient-noise pipeline under representation mismatch and
observing-run adaptation.  We reanalyse 10,429 detector--time strain candidates from 42 LIGO
O4a sessions with a frozen DINOv2 patch embedding and Top-$k$
multiple-instance score.  Candidate and native-background Q-transforms now
share $Q\in[4,64]$ and detector-specific thresholds are calibrated from 5,000
run-native windows by a non-overlapping temporal-block bootstrap.  The resulting statistical
dispositions contain 6,365 \robust, 1,275 \ambiguous, and 2,789 \background\
candidates. Detector-aware deduplication restores 57 windows omitted in the
preceding analysis;
among 10,372 paired historical candidates, 4,676 dispositions differ from the
preceding cross-representation analysis.

Direct controls resolve an O3b--O4a score shift and a reduction after native
adaptation for H1, but not for L1; O3b known-glitch separation is likewise
detector and morphology dependent.  The native-background draw gives
mean/minimum pairwise Spearman correlations of $0.674/0.515$ in a deliberately
near-boundary sample, while a final replicated study gives mean correlations
of $0.903$--$0.954$ near the boundary and $0.977$--$0.988$ in an
unconditioned population across background, clustering-seed, and dictionary-size
perturbations.
Whitening-context changes produce 37--40\% fixed-boundary flips and 63--70\%
after pad-specific recalibration in that boundary-conditioned sample.
Three-morphology, three-seed contamination tests demonstrate an
adaptation-absorption mechanism, and centred sine-Gaussian injections identify
a conditional low-$Q$ blind region.

Physical controls do not convert DSD survival into a discovery.  A conservative
H1--L1 max-shift screen yields 13/8,806 values above its threshold, but its
single on-source values and pooled per-event null maxima are not exchangeable;
the primary two-null PEM endpoint retains
2/26 \robust, 1/22 \ambiguous, and 7/93 \background\ events, with no resolved
\robust--\background\ enrichment ($p=1.000$); and two catalogue overlaps are
consistent with a circular-shift coverage proxy ($p=0.651$).
Simulation-only compact-binary controls show detector- and
distance-dependent disagreement between novelty flags, native dispositions,
and physical coincidence.  We therefore withdraw the preceding discovery, rate-limit,
catalogue-recall, and survey-wide stability interpretations.  The supported
result is a measured set of failure modes and validity conditions for
unsupervised detector characterization, not a new glitch class or an
astrophysical search.
\end{abstract}

\maketitle

\section{Introduction}
\label{sec:introduction}

Advanced LIGO, Advanced Virgo, and KAGRA form a network of increasingly
sensitive gravitational-wave (GW) observatories
\cite{LIGOScientific:2014pky,Acernese_2014,KAGRA:2020tym}.  Transient
non-Gaussian noise affects search sensitivity, background estimation, and
interpretation \cite{nuttall2018,davis2021,soni2025,pankow2018}.
Detector-characterization programmes
combine data-quality flags, environmental monitors, human expertise, and
machine learning to identify such disturbances.  Operational approaches also
include Omicron excess-power triggers, hierarchical auxiliary-channel vetoes,
and BayesWave signal--glitch models \cite{robinet2020,smith2011,cornish2015}.
Gravity Spy demonstrated the
value of supervised image classification and citizen science
\cite{biswas2013,powell2015,zevin2017,bahaadini2018,glanzer2023}, while iDQ models the relation between
strain artefacts and auxiliary degrees of freedom \cite{essick2020}.  Other
methods have used boosted networks \cite{mukund2017}, image transfer learning
\cite{razzano2018,george2018}, non-stationary noise regression
\cite{vajente2020,ormiston2020}, and multi-view O4 classifiers \cite{wu2024};
a broader review is given in \cite{cuoco2021}.

The complementary objective of discovering structure outside a fixed label set
has motivated similarity learning \cite{coughlin2019}, variational and
convolutional autoencoders \cite{sakai2022,laguarta2023}, and semi-supervised
anomaly searches such as GWAK \cite{raikman2023}.  These methods differ in
inputs and scientific target: some characterize glitches, some exploit
auxiliary channels, and some search for unmodelled astrophysical transients.
They share a central difficulty.  Detector noise is non-stationary, so distance
from a historical reference distribution can measure an observing-run change
rather than a rare transient.  A second difficulty follows when the reference
is adapted to the target run: genuine recurrent anomalies can enter the native
background model and be absorbed.

This paper studies those two failure modes empirically.  The Domain-Adaptive
Network for Transient Evaluation (DANTE) uses a frozen
DINOv2 ViT-S/14 encoder with registers \cite{oquab2024,darcet2024} on
constant-$Q$ spectrograms \cite{chatterji2004}.  No GW labels are used to train
the encoder.  An historical reference index supplies a primary novelty score;
a run-native index and block-bootstrap threshold form a Domain Shift Defense
(DSD).  A preceding DANTE study, published separately as
arXiv:2607.18136 \cite{dante_prior}, introduced the method and reported the O4a
candidate list.  The present reanalysis explicitly supersedes that study's
scientific interpretation: its object is not another candidate list but the
validity of the inference made from that list.

The contribution is fourfold.  First, we make the image representation,
background population, threshold population, and candidate query explicit and
machine-checkable.  Second, we propagate this contract through every
class-dependent experiment.  Third, we report negative controls and
sensitivity analyses even when they weaken earlier conclusions.  Fourth, we
separate statistical novelty, physical instrumental evidence, and
astrophysical response throughout.  This distinction is essential for the
broad GW context: an unsupervised anomaly flag can guide detector investigation
without being a detection statistic.

\begin{table*}[t]
\caption{Explicit transition from the principal preceding interpretations to
the claims supported in the present analysis. ``Not supported'' means that the
earlier inference is not used in the current conclusions; it does not imply that the underlying stored
score was fabricated.}
\label{tab:v5v6}
\small
\begin{ruledtabular}
\begin{tabular}{llll}
\vcell{3.0cm}{Preceding interpretation} &
\vcell{2.0cm}{Current status} &
\vcell{5.0cm}{Current evidence} &
\vcell{5.0cm}{Claim retained here} \\
\\[-1.5ex]\hline\\[-1.5ex]
\vcell{3.0cm}{Canonical stored scores are reproducible} &
\vcell{2.0cm}{Confirmed} &
\vcell{5.0cm}{The canonical pad-4 path reproduces 60/60 retained anchors to
maximum $|\Delta A|=6.85\times10^{-7}$.} &
\vcell{5.0cm}{The implementation anchor is reproduced before sensitivity
comparisons.} \\[1ex]
\vcell{3.0cm}{Q32 index queried by Q64 images is adequate} &
\vcell{2.0cm}{Corrected} &
\vcell{5.0cm}{A coherent Q64/Q64 contract changes 4,676 of 10,372 paired
historical dispositions; detector-aware deduplication restores 57 windows.} &
\vcell{5.0cm}{Index, threshold background, and queries must share a declared
representation.} \\[1ex]
\vcell{3.0cm}{The DSD taxonomy is physically meaningful} &
\vcell{2.0cm}{Restricted} &
\vcell{5.0cm}{The coherent taxonomy is 6,365/1,275/2,789, while physical
controls do not separate the labels.} &
\vcell{5.0cm}{\robust, \ambiguous, and \background\ are statistical
dispositions only.} \\[1ex]
\vcell{3.0cm}{The surviving singleton defines a novel glitch morphology} &
\vcell{2.0cm}{Not supported} &
\vcell{5.0cm}{The L1 window at catalogue GPS 1382955228 remains \robust\
under coherent rescoring, but the strain and PEM nulls provide no independent
counterpart or causal classification.} &
\vcell{5.0cm}{The event is an unclassified, statistically unusual L1-local
transient and is not evidence for a new glitch class.} \\[1ex]
\vcell{3.0cm}{Instrumental rate limits transfer unchanged} &
\vcell{2.0cm}{Not supported} &
\vcell{5.0cm}{The representation correction changes the class-dependent
funnel and its denominators.} &
\vcell{5.0cm}{No rate limit is reported in the present analysis.} \\[1ex]
\vcell{3.0cm}{Survivor ranking is invariant} &
\vcell{2.0cm}{Restricted} &
\vcell{5.0cm}{P5 gives mean/minimum $\rho=0.674/0.515$ over four draws; P4
gives $\rho=0.753$ at $K=512$.} &
\vcell{5.0cm}{Ranking is moderately stable in the tested near-boundary sample
and sensitive to dictionary size.} \\[1ex]
\vcell{3.0cm}{Whitening context changes at most about 5\%} &
\vcell{2.0cm}{Not supported} &
\vcell{5.0cm}{Fixed-boundary flips are 36.7--40.0\% and recalibrated flips are
63.3--70.0\% in a boundary-conditioned sample.} &
\vcell{5.0cm}{Preprocessing context is part of the calibrated model; rates are
not survey-wide.} \\[1ex]
\vcell{3.0cm}{Native adaptation only removes domain shift} &
\vcell{2.0cm}{Restricted + new control} &
\vcell{5.0cm}{Shift/adaptation are resolved for H1 but not L1; replicated
contamination suppresses three held-out morphologies.} &
\vcell{5.0cm}{Native adaptation has a measurable absorption failure mode in
the tested setup and is not universally required across detectors.} \\[1ex]
\vcell{3.0cm}{Known-glitch behaviour is implicit in novelty} &
\vcell{2.0cm}{New control} &
\vcell{5.0cm}{O3b Blip, Koi Fish, and Scattered Light comparisons show
detector- and morphology-dependent AUC.} &
\vcell{5.0cm}{These are binary construct controls, not multiclass recall.} \\[1ex]
\vcell{3.0cm}{\robust\ candidates are PEM-enriched} &
\vcell{2.0cm}{Not supported} &
\vcell{5.0cm}{The two-null endpoint gives 2/26 versus 7/93
($p=1.000$).} &
\vcell{5.0cm}{No auxiliary-coupling enrichment is resolved in the selected
public-channel sample.} \\[1ex]
\vcell{3.0cm}{Catalogue overlap measures recovery} &
\vcell{2.0cm}{Not supported} &
\vcell{5.0cm}{Two overlaps agree with a circular-shift coverage proxy
($p=0.651$).} &
\vcell{5.0cm}{The result is a null overlap test, not recall or detection
efficiency.} \\[1ex]
\vcell{3.0cm}{CBC controls establish signal safety} &
\vcell{2.0cm}{Restricted} &
\vcell{5.0cm}{The oracle-centred, simulation-only grid is detector-, source-,
and distance-dependent; BNS is unmeasured.} &
\vcell{5.0cm}{The injections characterize response under the tested protocol
only.} \\
\end{tabular}
\end{ruledtabular}
\end{table*}

\section{Data, scope, and prior knowledge}
\label{sec:data}

\subsection{O4a strain and analysis scope}

The analysis uses public H1 and L1 O4a strain from the Gravitational Wave Open
Science Center (GWOSC) \cite{gwosc_o4a_release,gwosc2023}.  Virgo is not included
because it did not participate in O4a science observations.  The configured
run bounds are GPS 1368975618--1389456018.  Forty-two macro-session identifiers
are processed for both LIGO detectors; the final catalogue contains candidates
from 40 H1 and all 42 L1 sessions.  The production log records ten unreadable
input blocks (nine H1 and one L1), and the historic scan predates an exact
successful-window ledger.  Calendar span is therefore not reported as searched
livetime.  The candidate
catalogue contains 10,429 unique detector--time windows after removing exact
same-detector repeats across overlapping sessions: 4,411 from H1 and 6,018
from L1.  Public CAT1 information is used
as the science-quality gate; the analysis is intentionally not a CAT2-vetoed
burst search because glitch-rich intervals are part of the detector-
characterization target.

The official O4a all-sky burst analysis calibrates coherent multi-detector
searches against explicit waveform families and astrophysical false-alarm
backgrounds \cite{lvk_burst_o4a}.  DANTE instead ranks detector-characterization
windows without assigning an astrophysical source model.  Its candidate counts,
nulls, and injection controls must therefore not be interpreted as burst-search
significance, efficiency, or rate limits.

The workflow is retrospective.  Its native index is built from the run under
study, excluding candidate-adjacent intervals, and must be recalibrated for a
new run.  The software's discovery/aggregation contract is exercised
end-to-end for O4a and O3b, but the scientific results in this article are O4a
only.  We make no claim of low-latency readiness or of validation on a future
O5 run.

\subsection{Three distinct populations}

The primary reference is an O3b vector-quantized dictionary used to flag
windows that are distant from historical labelled morphology.  It contains
$K=275$ centroids in the production artefact.  The DSD uses a separate O4a
native index with $K=1216$, built from 1,294 vetoed clean background segments
balanced across H1 and L1.  A seeded sample of 50,000 pre-quantization tokens
is retained for reproducibility and point-to-point controls.  Finally, the DSD
thresholds are calibrated on independent chronological background samples of
5,000 windows per detector.  Index construction, threshold calibration, and
candidate evaluation are therefore separate populations.

\section{Representation-coherent pipeline}
\label{sec:methods}

\subsection{Preprocessing and patch representation}

Each candidate is represented by a 32\,s analysis window at 4096\,Hz.  Strain is fetched
with 4\,s of context on each side, whitened on the padded interval, cropped back
to the clean subwindow, and band-passed once.  At this sample rate the configured
20--2000\,Hz passband is capped at 1843.2\,Hz.  The production DSD Q-transform
requests $Q\in[4,64]$ and 20--2048\,Hz; GWpy lowers the effective logarithmic
axis to 20--1291.05\,Hz for this window and Q range.  Its time-by-frequency
array is resized without transposition to a $256\times256$ cividis image.  The
frozen DINOv2 ViT-S/14 encoder maps the image
to 384-dimensional patch tokens.  The pipeline uses patch tokens, not the
global class token, because a short transient may occupy only a small part of
the image.

For normalized patch token $\mathbf z_j$ and normalized reference centroids
$\mathbf c_\ell$, the patch distance is
\begin{equation}
 d_j = \min_\ell \left(1-\mathbf z_j^\mathsf{T}\mathbf c_\ell\right).
\end{equation}
Let $\mathcal T_k$ denote the $k=68$ largest patch distances.  The
multiple-instance anomaly score is
\begin{equation}
 A(\mathbf x)=\frac{1}{k}\sum_{j\in\mathcal T_k}d_j .
\label{eq:mil}
\end{equation}
All saliency maps and stored MIL vectors use these same selected patches.

\subsection{Coherent native calibration}

Earlier catalogues queried a $Q_{\max}=32$ native index with
$Q_{\max}=64$ candidate images.  That comparison is retained only as a legacy
cross-representation result.  The present analysis rebuilds the native index
at $Q\in[4,64]$ and re-scores every candidate using the identical
representation.  The index path, SHA256, Q-range, embedding dimension, and
normalization contract are stored with each dependent artefact.  All 10,429
candidates completed coherent scoring; none was silently dropped. Exactly
10,372 scores were reused after detector--GPS and hash verification and 57
were newly computed. Within the paired historical population, the
representation change alters 4,676 dispositions, showing that this was a
scientific correction rather than metadata hygiene.

For each detector, chronological background scores are resampled in temporal
blocks.  Repeating the non-overlapping-block bootstrap yields a confidence interval
$[\tau_{\rm lo},\tau_{\rm hi}]$ for the background $P_{99}$.  We define
\begin{align}
 A &> \tau_{\rm hi} &&\Rightarrow \robust,\\
 \tau_{\rm lo}\le A\le\tau_{\rm hi} &&\Rightarrow \ambiguous,\\
 A &< \tau_{\rm lo} &&\Rightarrow \background.
\end{align}
For H1, $(P_{99},\tau_{\rm lo},\tau_{\rm hi})
=(0.16158,0.13023,0.20409)$; for L1 the values are
$(0.17606,0.15032,0.21959)$.  Repeating the complete $B=10^6$ calibration ten
times gives endpoint Monte Carlo standard deviations no larger than
$4.89\times10^{-5}$; the H1 upper and L1 lower endpoints are identical in all
ten repeats.  The largest Monte Carlo standard deviation is
$7.1\times10^{-4}$ of its confidence-interval width.

The production rule $b=\lfloor n^{1/3}\rfloor=17$ is also varied over
$b\in\{8,17,32,64\}$ with 200,000 replicas.  Non-overlapping blocks change
46, 1, 42, and 140 of 10,429 dispositions relative to production; an
overlapping moving-block control changes 86, 43, 7, and 148.  The maximum is
1.42\% of the catalogue and is concentrated at the calibrated boundaries, so
the global conclusion is stable but individual boundary labels retain a
block-model dependence.

\subsection{Physical and catalogue nulls}

Cross-detector coincidence is evaluated in whitened strain, not in embedding
space.  We maximize the normalized H1--L1 cross-correlation over the
inter-site light-travel interval plus a 2\,ms margin.  For each event we retain
the maximum over 4--8 valid time shifts; pooling these per-event maxima gives
$P_{99}$ threshold $\tau_{\rm cc}=0.4046$.  Among 8,806 evaluable candidate
windows, 13 single on-source values exceed this threshold.  Because an
on-source statistic is a single value whereas its null contribution is a
per-event maximum, this conservative screen is not an exchangeable tail-count
test and the 13/8,806 count is not assigned a $p$-value or a physical
interpretation.  The remaining 1,623 historical candidates have no validated
measurement and their missingness is not assumed random.

For environmental coupling, magnitude-squared coherence is calculated with
Welch's method \cite{welch1967} across the public auxiliary channels available
for each event.  The primary endpoint requires the observed family-wise maximum
to exceed both an event-specific time-shift $P_{99}$ and a quiet-background
zero-lag $P_{99}$.  This second null is necessary because persistent spectral
coherence can make time shifts alone non-discriminating.  Channels in the
public release are not treated as safety-certified veto channels.

The event-specific time-shift null uses a nearby 4\,h CAT1-clean block with a
96\,s exclusion around every candidate.  Coherence is measured over
20--500\,Hz with 2\,s Welch segments and 1\,s overlap.  The null uses ordered
pairs of 32\,s windows on a 96\,s stride separated by at least 64\,s; one common
shift is applied to all available channels, preserving their dependence.  The
event-level maximum over channel and frequency is calibrated at family-wise
$\alpha=0.01$, with threshold uncertainty resampled over window indices.  The
second threshold is the quiet-window zero-lag $P_{99}$.

Catalogue overlap uses GWTC-4.0/4.1 event times falling in the O4a interval.
Because the historical scan did not persist an exact successful-window ledger,
raw block coverage is an upper-bound proxy.  The null circularly shifts the
complete catalogue by 10,000 common offsets of at least one day, preserving
its internal temporal structure.

Unless otherwise stated, uncertainty intervals are two-sided 95\% intervals.
The DSD uses $10^6$ complete-block resamples; cross-run differences and
known-glitch AUCs use 2,000 resamples within their independent groups, paired
adaptation resamples paired score differences, and final robustness intervals
use 1,000 candidate resamples of each fixed score matrix.  Absorption $z$
intervals independently resample injected and background scores 2,000 times;
binomial fractions use Wilson intervals and PEM association uses a two-sided
Fisher exact test.  Only the coherent DSD disposition and the declared strain
and PEM nulls are primary endpoints.  The remaining tests are diagnostic, no
no global multiplicity adjustment is applied across them, and they do not support
discovery claims.

\section{Results}
\label{sec:results}

\subsection{Detector-dependent domain shift and known-glitch controls}

We test the premise for native adaptation directly before interpreting the O4a
taxonomy.  For each detector, 100 vetoed clean windows from O3b and 100 from
O4a are split chronologically into 60 dictionary-training and 40 held-out
windows.  We set the dictionary size by the same compression-density rule used
for the production native index: each training pool contains $60\times1369$
patch tokens, and 1,458 tokens per centroid gives matched $K=56$ O3b and O4a
dictionaries after rounding.  The smaller absolute $K$ therefore tracks the
smaller training pool and is not, by itself, evidence of stronger compression.
This construct-control size is nevertheless distinct from the production O3b
($K=275$) and native O4a ($K=1216$) dictionaries, so the score-based cross-run
endpoints remain conditional on these samples and the $K=56$ control.  With the fixed O3b control dictionary, the H1
O4a-minus-O3b mean score
difference is $0.03983$ (95\% bootstrap interval $[0.02914,0.05184]$;
two-sample KS $p=1.23\times10^{-7}$).  Replacing that dictionary with its
O4a-native counterpart lowers the paired H1 O4a score by $0.02737$ on average
($[-0.03296,-0.02158]$).  A five-block out-of-fold run probe, computed from
L2-normalized segment-mean embeddings without either $K=56$ dictionary, gives
AUC $0.920$ ($[0.881,0.955]$).

The corresponding L1 endpoints do not resolve the same effect: the
O4a-minus-O3b difference is $-0.00290$ ($[-0.01304,0.00823]$; KS $p=0.097$),
the native-minus-cross-index change is $0.00035$
($[-0.00513,0.00582]$), and run-probe AUC is $0.616$
($[0.535,0.694]$).  Native adaptation is therefore detector-dependent in
these samples, rather than a universal correction for O4a.

\begin{figure*}[t]
\centering
\includegraphics[width=0.94\textwidth]{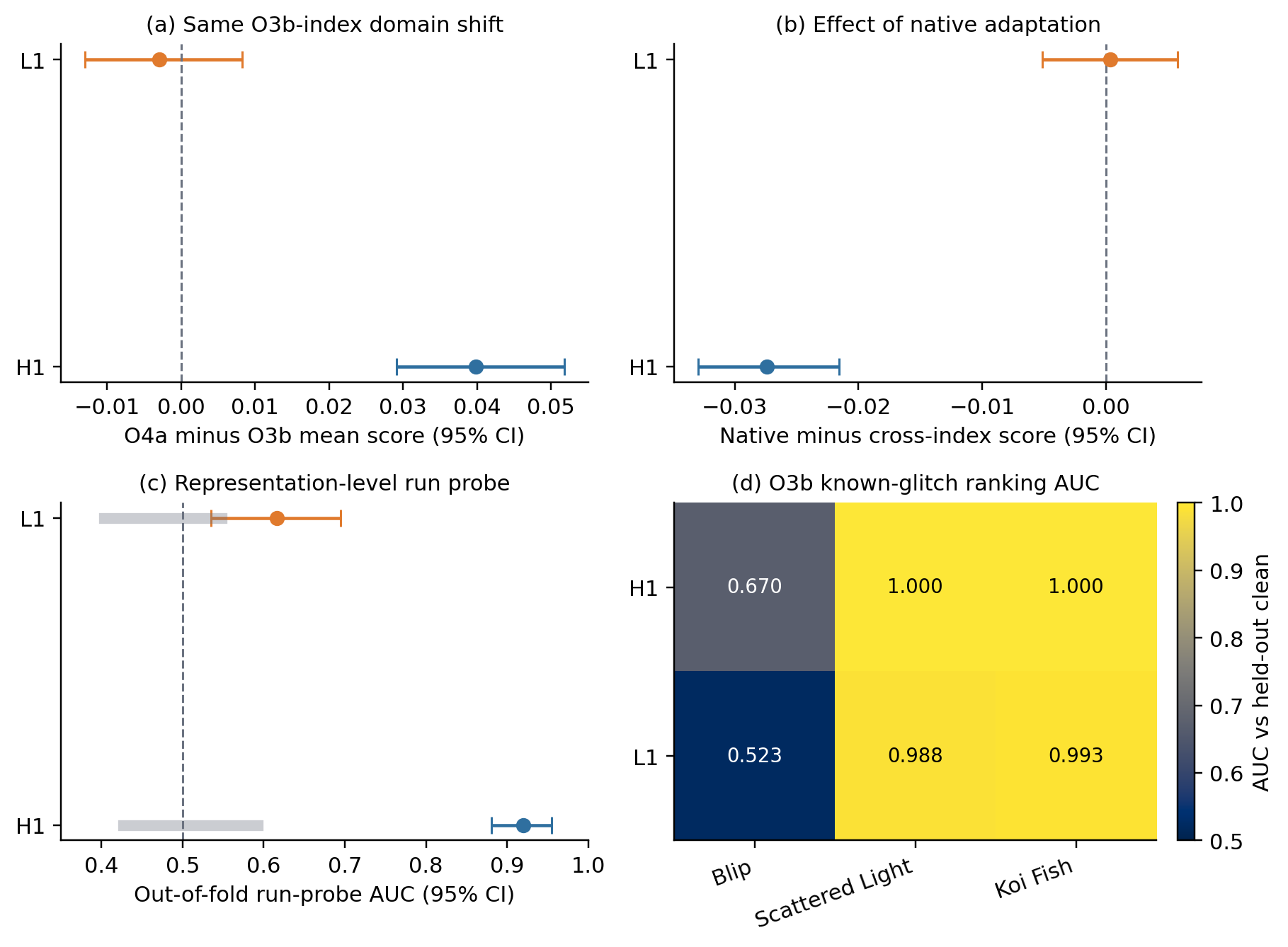}
\caption{Direct domain and known-glitch controls under the coherent Q64
representation.  Panels show held-out cross-run score shift, paired score
change after native adaptation, out-of-fold run-probe AUC, and binary AUC for
three labelled O3b glitch morphologies against held-out clean strain.  Error
bars are 95\% bootstrap intervals.}
\label{fig:domain-known}
\end{figure*}

As an external construct control, the same O3b dictionaries score 30
Gravity-Spy-labelled examples \cite{gspy_zenodo} of each of Blip, Scattered
Light, and Koi Fish per detector against 40 held-out clean O3b windows.  AUCs
are $0.670$, $1.000$, and $1.000$ for H1, and $0.523$, $0.988$, and $0.993$
for L1.  These binary, morphology-specific controls show weak L1 Blip
separation and are not O4a multiclass recall (Fig.~\ref{fig:domain-known}).

\subsection{Discovery disposition is not physical classification}

\begin{figure}[t]
\centering
\includegraphics[width=\columnwidth]{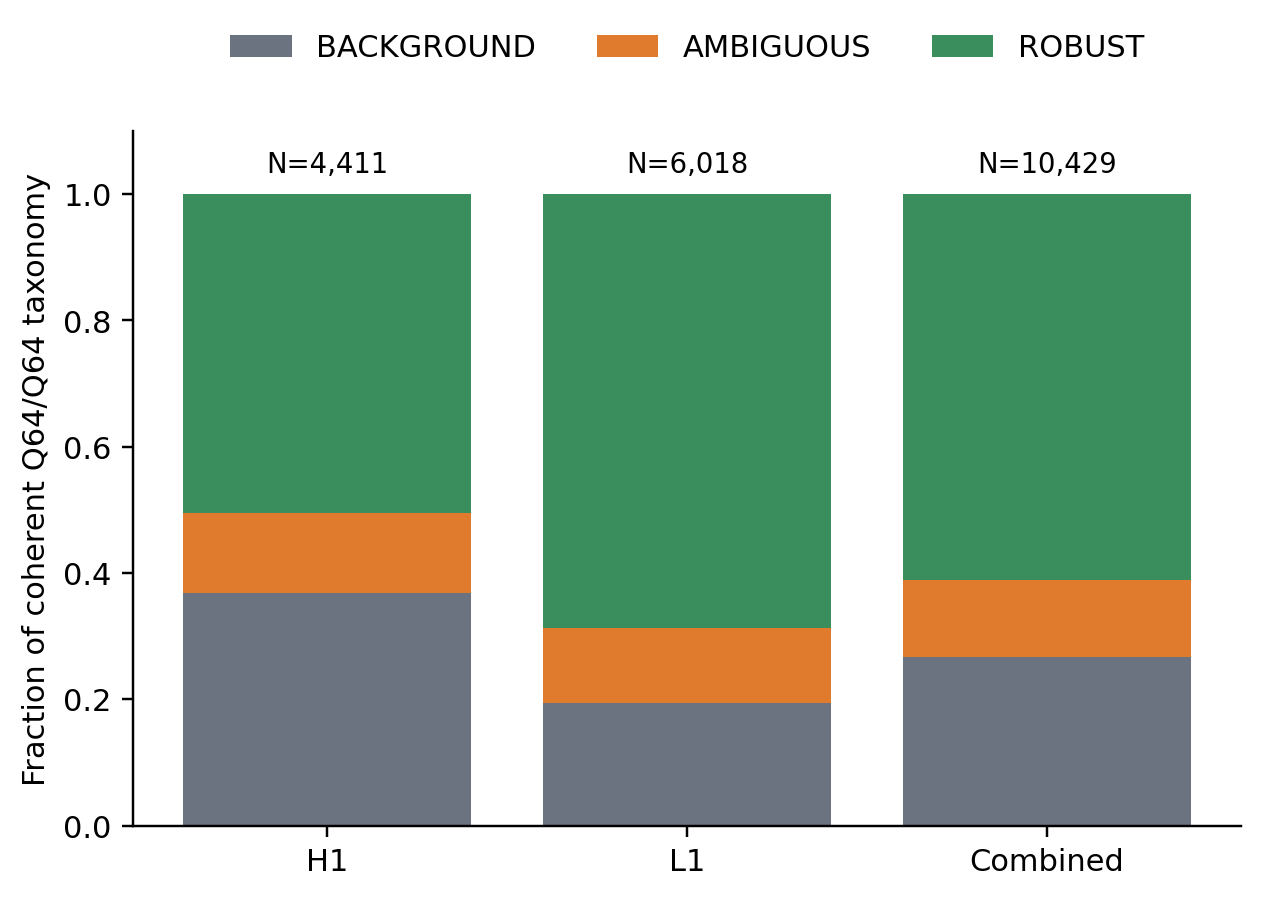}
\caption{Coherent Q64/Q64 DSD disposition by detector and combined.  Counts
above each bar include every evaluated candidate.}
\label{fig:funnel}
\end{figure}

The coherent taxonomy contains 6,365 \robust, 1,275 \ambiguous, and 2,789
\background\ windows (Fig.~\ref{fig:funnel}).  Detector-specific \robust\
fractions are 50.5\% for H1 and 68.8\% for L1.  These labels state where a
score lies relative to a run-native background interval.  They do not by
themselves identify a glitch family, an instrumental cause, or an
astrophysical signal.

At the graph-distance cut $D_{\rm cut}=0.25$, fixed before the final execution, the topology is
saturated.  The largest connected component contains 99.92\% of \robust\
candidates and 100\% of \ambiguous, \background, and 3,000 unselected
native-background vectors.
Identical clustering therefore does not distinguish anomaly status.  Likewise,
the strongest cross-session pairs occur at approximately the all-pairs
cross-session rate, and the mean neighbour-session span is below a shuffled
session null.  The current embedding resolves no positive recurrence signal;
this is not evidence that no physical morphology recurs.

\subsection{Sensitivity of the statistical disposition}

\begin{figure*}[t]
\centering
\includegraphics[width=0.92\textwidth]{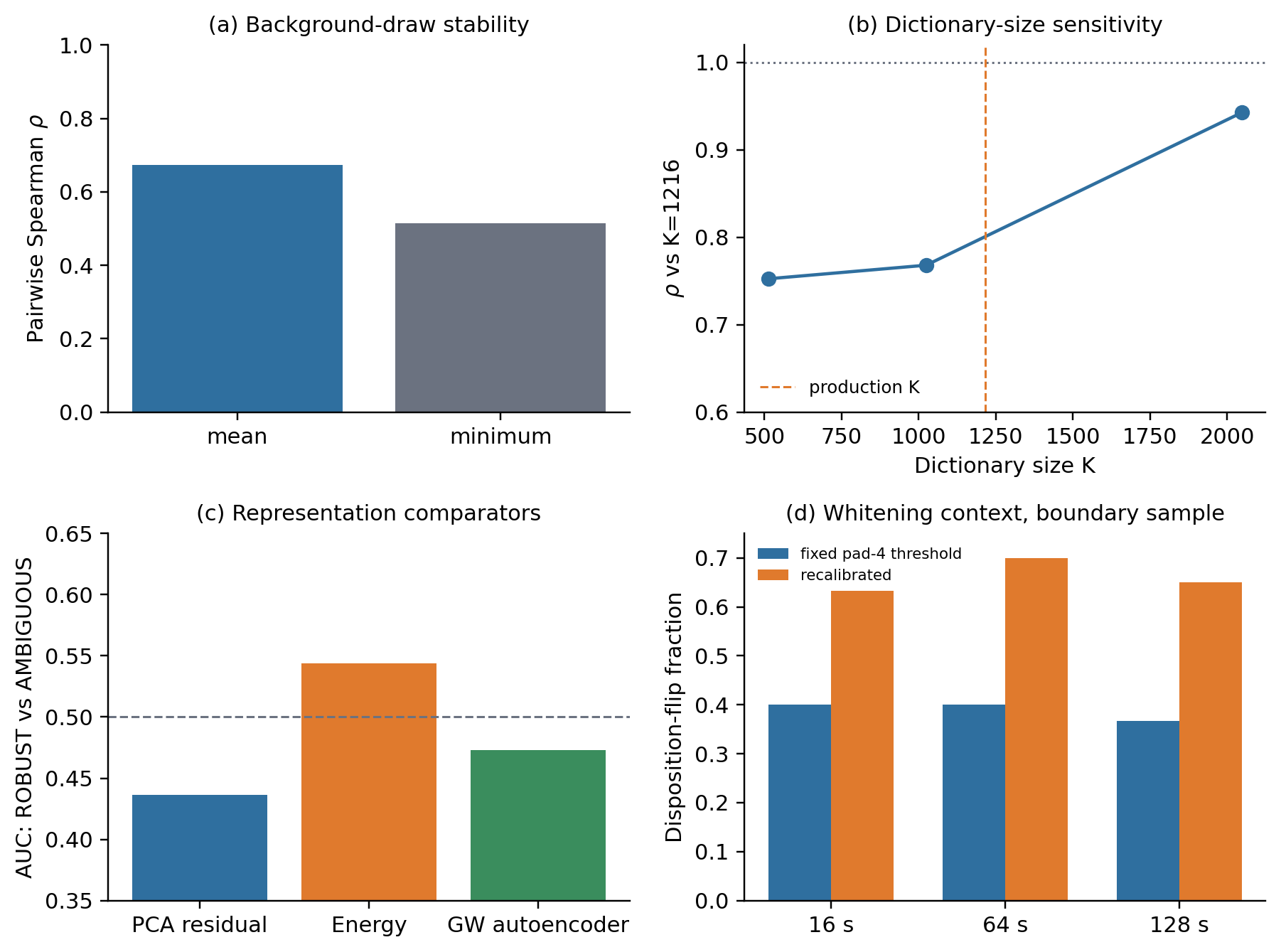}
\caption{Sensitivity and baseline controls on representation-coherent,
near-boundary samples.  (a) Stability to native-background draw.  (b) Rank
correlation versus the production dictionary size.  (c) Classical pixel
baselines and a GW-specific convolutional-autoencoder negative control.
(d) Candidate disposition under longer whitening contexts.}
\label{fig:robustness}
\end{figure*}

Four independent 1,300-segment native-background draws give mean pairwise
Spearman correlation $\rho=0.674$ and minimum $0.515$ for a deliberately
near-boundary sample of 80 \robust\ and 80 \ambiguous\ candidates
(Fig.~\ref{fig:robustness}a).  The median per-candidate score standard
deviation is 0.00757.  Candidate ordering is positively, but only moderately,
correlated across the four background draws in this selected sample.

Dictionary size also matters (Fig.~\ref{fig:robustness}b).  Relative to
$K=1216$, rank correlation is 0.753, 0.768, and 0.943 for
$K=512,1024,2048$, respectively.  Ranking becomes more similar near and above
the production size but is not invariant to $K$.  As simple alternatives, a
PCA reconstruction residual and total spectral energy yield AUC 0.436 and
0.544 for separating the same coherent classes, and correlations $-0.158$ and
$0.030$ with DANTE (Fig.~\ref{fig:robustness}c).  Energy AUC is
detector-dependent (0.598 for H1 and 0.490 for L1).  A detector-specific
convolutional autoencoder trained from scratch on 650 candidate-vetoed O4a
Q-transform backgrounds per detector is also non-discriminating: across three
training seeds its pooled AUC is 0.473 (seed range 0.473--0.485) and its
Spearman correlation with DANTE is $-0.047$ (H1/L1 AUC 0.528/0.424).
This negative control shows that nonlinear GW-specific reconstruction error
does not reproduce the DANTE disposition; none of these comparators
establishes the optimality or causal morphological meaning of natural-image
features.

The canonical 4\,s whitening-context implementation reproduces all 60 retained
anchor scores to a maximum absolute difference $6.85\times10^{-7}$.  Six of
66 attempted candidates are excluded by a hashed ledger for non-finite context
or anchor mismatch.  The retained near-boundary disposition is sensitive to
context (Fig.~\ref{fig:robustness}d).  Relative to pad 4\,s, fixed-threshold
flip fractions are 40.0\%, 40.0\%, and 36.7\% at pads 16, 64, and 128\,s;
after recalibrating 5,000 background windows per detector at every pad they are
63.3\%, 70.0\%, and 65.0\%. The median score swing is 0.0109 and 16/60 exceed
0.02. These are boundary-conditioned sensitivities, not
survey-wide failure rates.

\begin{figure*}[t]
\centering
\includegraphics[width=0.92\textwidth]{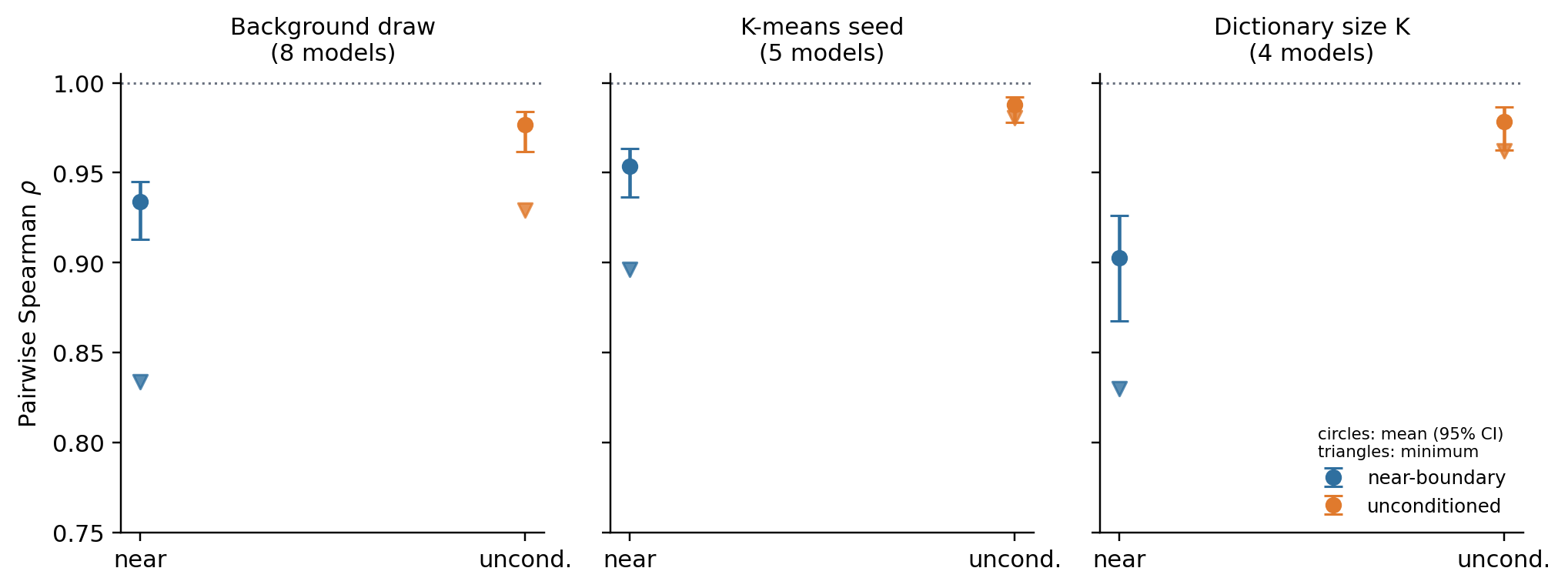}
\caption{Final replicated robustness study.  Circles show mean pairwise
Spearman correlation with 95\% candidate-bootstrap intervals; triangles show
the minimum pair.  Near-boundary and unconditioned pools are reported
separately for background draw, clustering seed, and dictionary size.}
\label{fig:robustness-replicates}
\end{figure*}

A separate final replication holds the candidate-token pools fixed and varies
one component at a time (Fig.~\ref{fig:robustness-replicates}).  Across eight
background draws, mean pairwise rank correlation is $0.934$
($[0.913,0.945]$; minimum pair $0.834$) near the boundary and $0.977$
($[0.962,0.984]$; minimum $0.929$) in an unconditioned 160-candidate pool.
Across five clustering seeds the corresponding means are $0.954$
($[0.936,0.963]$; minimum $0.896$) and $0.988$
($[0.978,0.992]$; minimum $0.981$).  Across
$K\in\{512,1024,1216,2048\}$ they are $0.903$
($[0.868,0.926]$; minimum $0.830$) and $0.978$
($[0.963,0.987]$; minimum $0.962$).  The unconditioned population is
consistently more stable; it does not replace the boundary-conditioned C2
endpoint because the populations and perturbation ensembles differ.

\subsection{Native adaptation can absorb recurrent anomalies}

We inject controlled Blip, Koi Fish, and Scattered Light morphologies at fixed
peak amplitude 12 in whitened-strain units and durations 1, 1, and 1.5\,s,
respectively, into fractions
$\{0,0.02,0.05,0.10,0.15,0.20,0.30,0.40\}$ of a 300-segment O4a L1
native-index training set and evaluate 60 held-out injections.  Scaling the
production tokens-per-centroid ratio gives $K=282$, distinct from the
production $K=1216$ index.  Each morphology is repeated with three clustering
seeds.  An absorption crossing fixed before the final replicated execution
requires both standardized
injection/background separation $z\le3$ and a held-out fraction above the
index-specific background $P_{99}$ no larger than 0.5.

The generator normalizes each waveform to unit peak before scaling and adds it
at a seeded, edge-protected random position after whitening.  Amplitude 12 is
therefore an SNR-like whitened-noise scale, not matched-filter SNR.  The held-out
60 injected and 150 background segments are disjoint from index construction;
a same-size all-background dictionary is the composition control.

\begin{figure*}[t]
\centering
\includegraphics[width=0.94\textwidth]{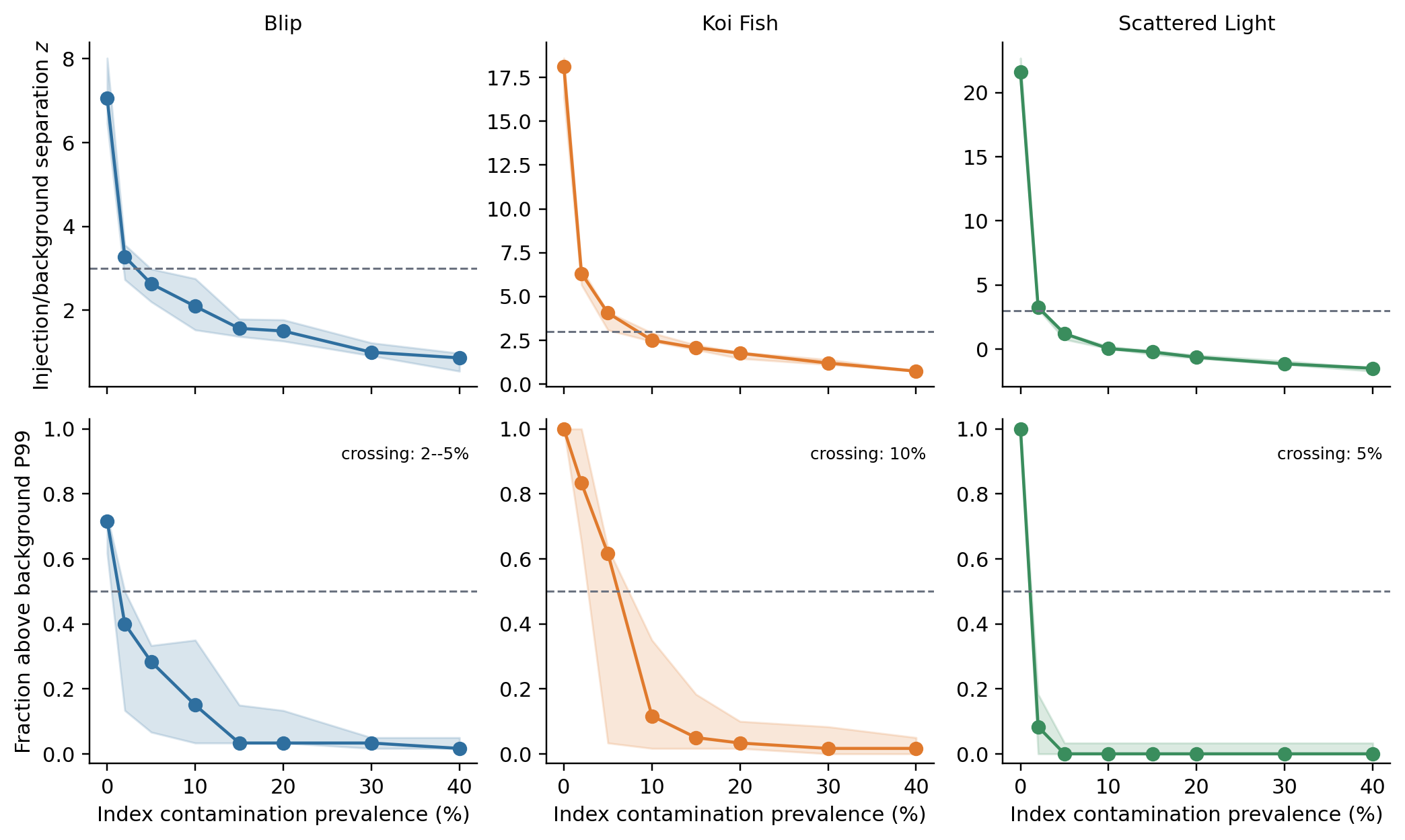}
\caption{Controlled native-index absorption for three injected morphologies
and three clustering seeds.  Lines show across-seed medians and bands show the
seed range.  Dashed lines mark the joint crossing rule.  Experimental
$K=282$ dictionaries are distinct from the production $K=1216$ index.}
\label{fig:absorption-matrix}
\end{figure*}

All seeds cross the rule: Blip at 2--5\% prevalence, Koi Fish at 10\%, and
Scattered Light at 5\% (Fig.~\ref{fig:absorption-matrix}).  At zero
contamination, median $z$ is 7.05, 18.13, and 21.61, while the median flagged
fraction is 0.717, 1.000, and 1.000.  Same-size all-background controls remain
strongly separated.  The mechanism is reproduced across these synthetic
morphologies and seeds, but crossings remain conditional on detector,
amplitude, duration, sampling, representation, and rule.

\subsection{Empirical blind region}

\begin{figure}[t]
\centering
\includegraphics[width=\columnwidth]{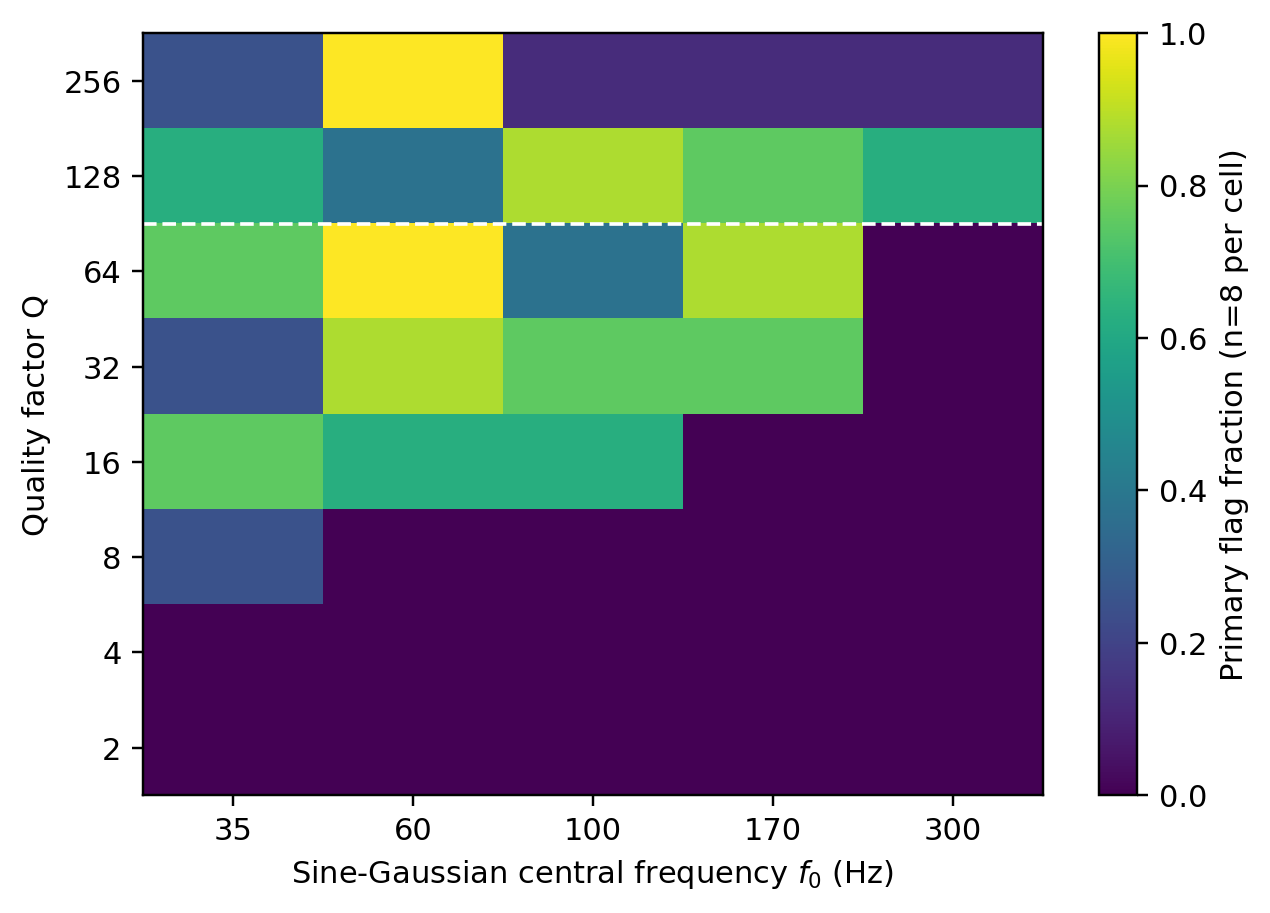}
\caption{Primary flag fraction for centred sine-Gaussian injections at fixed
SNR 20.  The dashed line separates $Q\le64$ from $Q>64$; the dominant blind
region is at low Q, not uniformly above the representation boundary.}
\label{fig:blind}
\end{figure}

A centred sine-Gaussian grid spans five frequencies, eight Q values, and eight
realizations per cell at fixed injected matched-filter SNR 20
(Fig.~\ref{fig:blind}).  Each waveform is scaled with
$\rho^2=4\int_{20\,\mathrm{Hz}}^{2000\,\mathrm{Hz}}
|\tilde h(f)|^2/S_n(f)\,\mathrm df$, using the clean 32\,s segment and 4\,s
PSDs, then centred before canonical preprocessing.  Mean
primary flag rate is 26.3\% for $Q\le64$ and 48.8\% for $Q>64$.  Every
$Q=2$ and $Q=4$ cell has zero recovery and most $Q=8$ cells also fail.  The
tested grid does not support a universal loss of sensitivity above
$Q_{\max}=64$; under this fixed-SNR protocol it instead shows a broad low-Q
blind region with additional frequency-dependent structure.  Individual-cell
binomial intervals are wide because $n=8$.

\subsection{Physical controls}

\begin{figure}[t]
\centering
\includegraphics[width=\columnwidth]{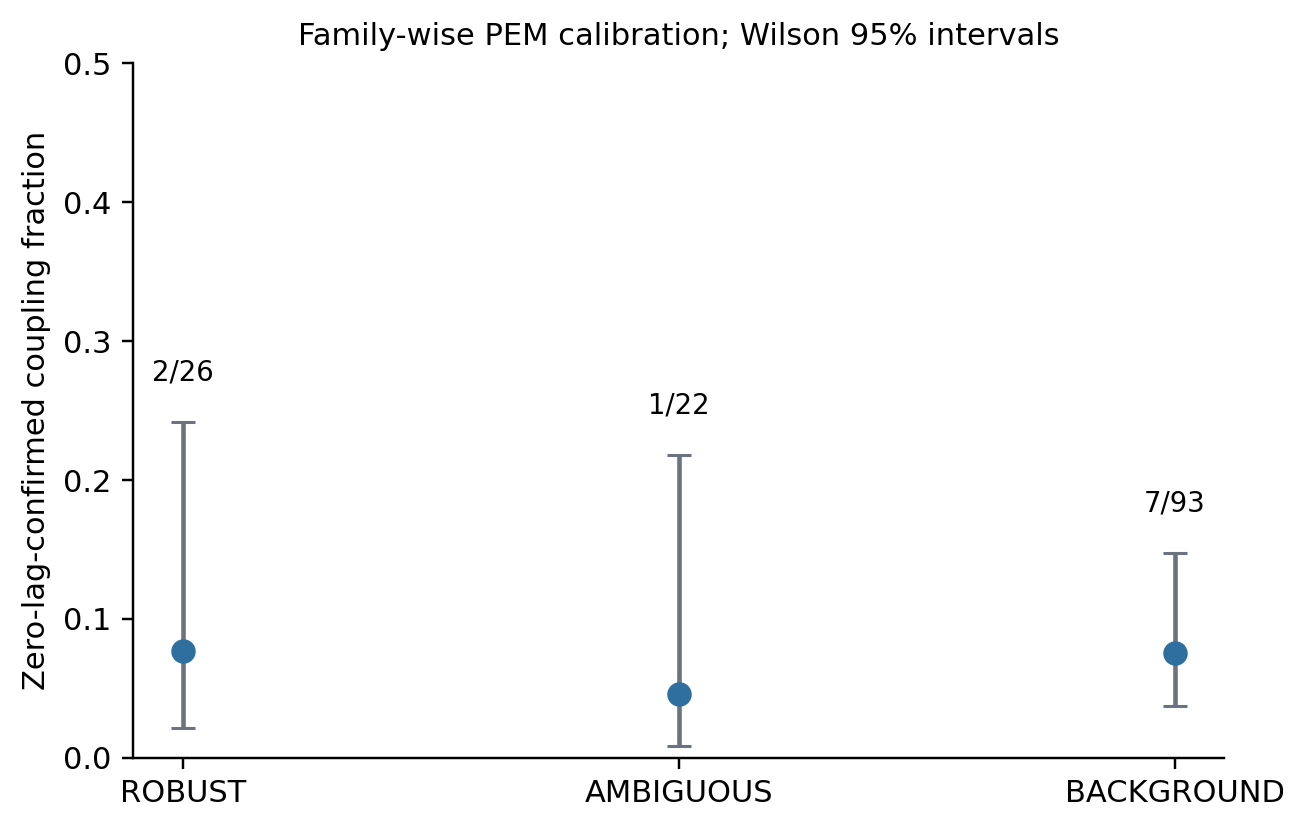}
\caption{Primary family-wise PEM endpoint with Wilson 95\% intervals.
All 141 selected events are calibrated; the intervals remain broad for the
smaller coherent classes.}
\label{fig:pem}
\end{figure}

The PEM experiment uses a fixed measured cohort of 141 events originally
selected by spacing over score rank within the already coherent Q64/Q64
taxonomy.  Exact detector--GPS rejoining after a subsequent update of that
coherent taxonomy changes eight class labels and gives 26 \robust, 22
\ambiguous, and 93 \background\ events; this is not a newly drawn
class-balanced sample.  All 141 receive an empirical family-wise calibration.
A time-shift threshold alone is not a coupling
endpoint: 6/26 \robust\ and 34/93 \background\ events exceed it (odds ratio
0.521, two-sided Fisher $p=0.245$), while 38 of the 48 exceedances fail the
quiet-background zero-lag control.  The primary endpoint retains 2/26
\robust, 1/22 \ambiguous, and 7/93 \background\ events
(Fig.~\ref{fig:pem}); for \robust\ versus \background\ the odds ratio is
1.02 with two-sided Fisher $p=1.000$.  Thus the coherent sample does not
resolve class enrichment in auxiliary coupling.  The point estimate is not
evidence of equal rates, and the public-channel inventory does not exclude
unmeasured couplings.

\subsection{Reassessment of the preceding singleton}

\begin{figure*}[t]
\centering
\includegraphics[width=0.96\textwidth]{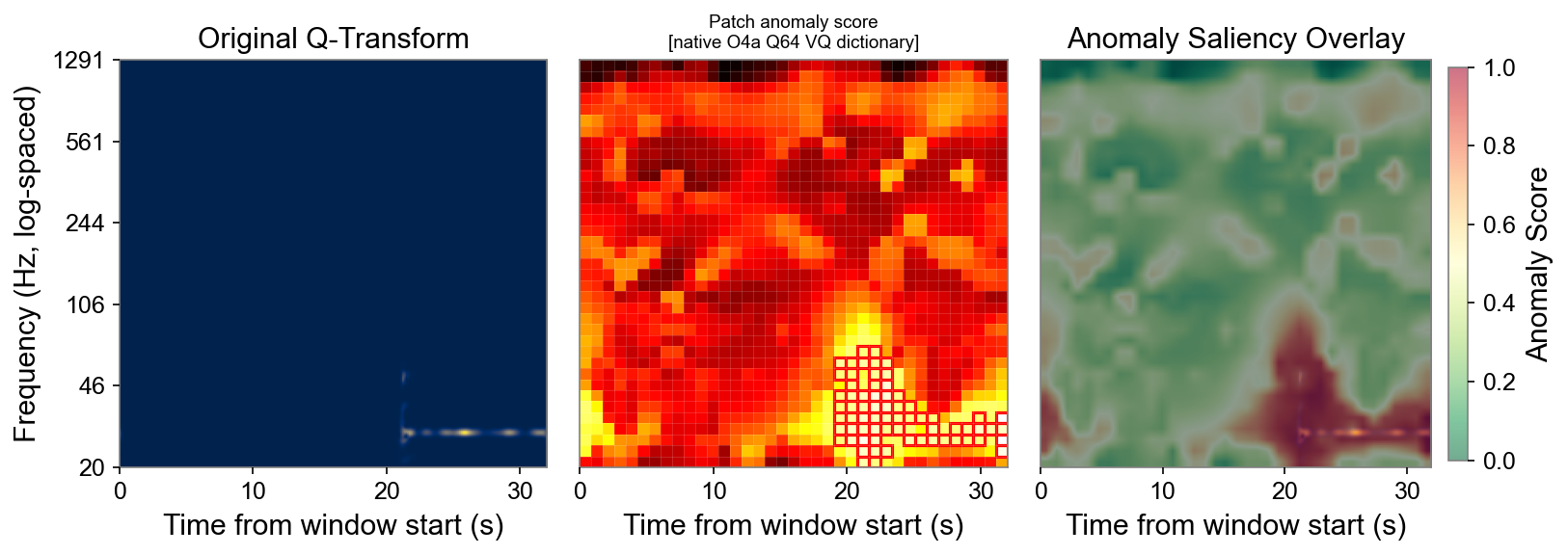}
\caption{Coherent-Q64 reassessment of the preceding L1 singleton.  The left panel is
the normalized Q-transform of the exact 32 s analysis window.  The centre
panel shows patch distances to the frozen native O4a dictionary; red boxes are
the same Top-68 patches pooled into the stored score.  The overlay is a
visualization of this statistical score, not a morphology or causal label.}
\label{fig:candidate}
\end{figure*}

The L1 entry at catalogue GPS 1382955228 labels the padded historical crop;
the interval actually scored is GPS 1382955232--1382955264.  The coherent
native score is 0.598877, above the L1 \robust\ threshold, and the production
Top-68 patches in Fig.~\ref{fig:candidate} reproduce that stored score to
$1.2\times10^{-7}$.  An independent descriptive implementation adapted from
a public proposal by Kretski and archived with the versioned analysis code and
evidence bundle \cite{dante_zenodo,dante_v6_evidence}
places the ringing feature at GPS 1382955253.17, near 28 Hz, with in-band
energy 304 times the mean of 16 adjacent windows.  These descriptors establish
that the transient is loud and localized in L1; they do not establish its
physical origin.

The event exceeds its scale-specific $P_{99}$ threshold at 0.5, 1, 2, and
4 s, with the largest margin (0.112) at 4 s.  Its on-source H1--L1
coincidence coefficient is 0.0716, below both the time-shift mean 0.197 and
maximum 0.286; patch intersection-over-union is 0.0074.  The public-channel
PEM statistic is 0.478, below the family-wise threshold 0.663 and the quiet
zero-lag threshold 0.789.  Thus no H1 counterpart or coupling in the tested
public auxiliary subset is resolved.  The feature is qualitatively compatible
with low-frequency scattered-light-like ringing, but remains unclassified:
the present analysis does not call it a new morphology, and the limited PEM inventory cannot
exclude an unmeasured instrumental coupling.

The catalogue control covers 131 of 135 O4a catalogue events in at least one
detector under the raw-block proxy and 118 in both.  Two H1 candidate windows
overlap a catalogue event and none overlap in both detectors.  Circular shifts
give $2.190\pm1.467$ expected any-detector overlaps, a 95\% interval $[0,5]$,
and empirical $p=0.651$.  This is no resolved overlap excess.  It is not a
recall measurement because exact historical processed-window coverage is
unavailable.

\begin{figure}[t]
\centering
\includegraphics[width=\columnwidth]{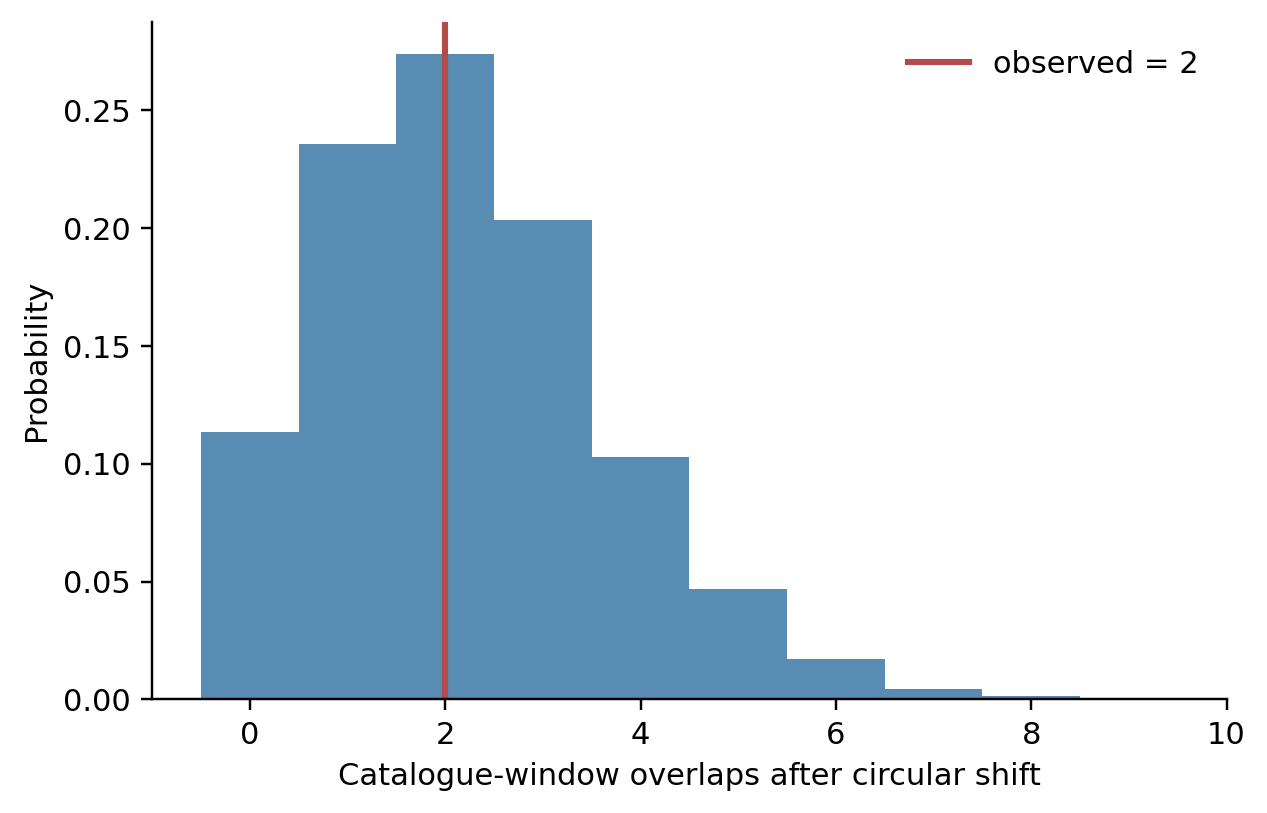}
\caption{Circular-shift coverage proxy for catalogue overlap.  The observed
two any-detector overlaps lie inside the null distribution; this is not a
catalogue-recall measurement.}
\label{fig:catalog-null}
\end{figure}

\subsection{Compact-binary injection controls}

\begin{figure*}[t]
\centering
\includegraphics[width=0.94\textwidth]{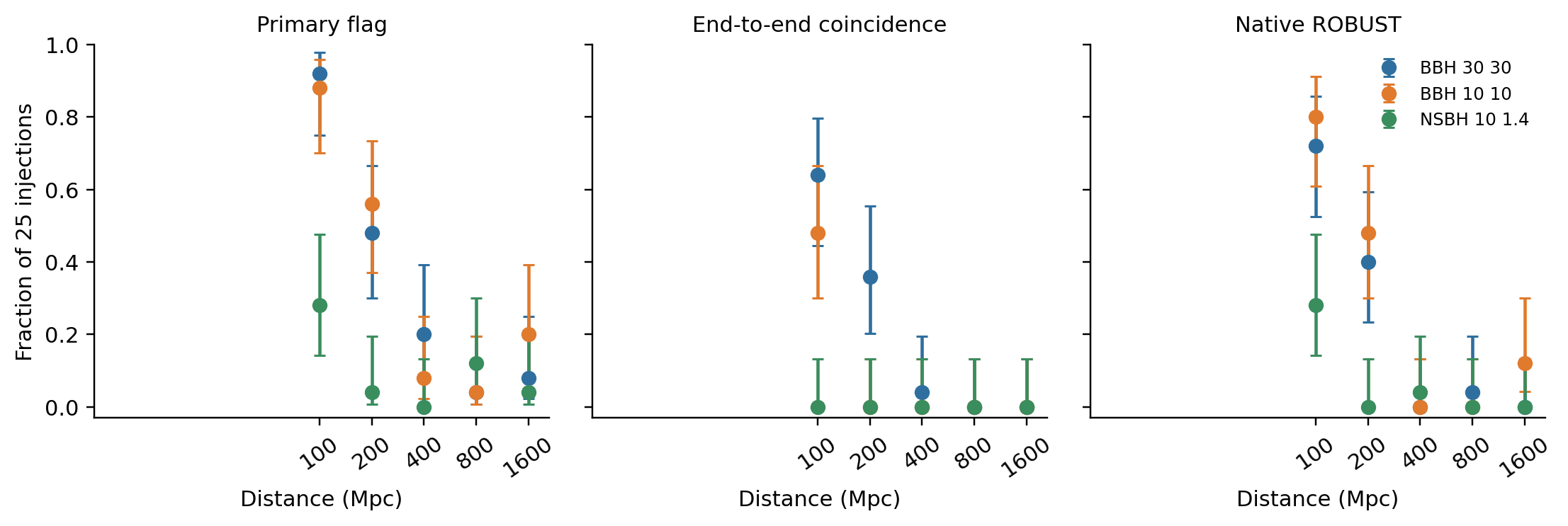}
\caption{Simulation-only compact-binary controls.  Points show the primary
flag, end-to-end physical coincidence, and maximum single-detector
\robust\ fraction; binomial error bars are Wilson 95\% intervals.  Each point
contains 25 successful injections into real O4a noise.}
\label{fig:cbc}
\end{figure*}

We inject zero-spin, non-precessing IMRPhenomD waveforms into raw real vetoed
O4a noise for BBH $30+30\,M_\odot$, BBH $10+10\,M_\odot$, and NSBH
$10+1.4\,M_\odot$ at 100, 200, 400, 800, and 1,600\,Mpc.  With seed 42, sky
position is isotropic, polarization is uniform, inclination is uniform in
$\cos\iota$, and H1/L1 use their own antenna responses and delays.  Lower
frequencies are 20, 25, and 30\,Hz; the merger is centred and paired clean
segments are retained.  A successful trial requires finite data in both
detectors and completion of the full scoring/coincidence path; invalid fetches
are skipped rather than counted as misses.  There are 25 successful trials per
cell (375 injections).  For $30+30\,M_\odot$ at 100 and 200\,Mpc, the primary score
flags 23/25 and 12/25 trials, while end-to-end physical coincidence recovers
16/25 (Wilson 95\% interval 0.445--0.798) and 9/25
(0.202--0.555).  For $10+10\,M_\odot$, primary flagging is 22/25 and 14/25,
but coincidence is 12/25 and 0/25.  NSBH coincidence is zero in all tested
cells.  Native DSD disposition is detector- and distance-dependent rather than
a monotonic continuation of the primary score (Fig.~\ref{fig:cbc}).

The simulation demonstrates morphology-dependent response and a tension
between historical novelty and native adaptation.  It is not an end-to-end
astrophysical efficiency: merger placement is known by construction, only one
zero-spin intrinsic point is tested per named system, and BNS
$1.4+1.4\,M_\odot$ is not measured
because its 31\,s waveform violates the current 32\,s safety margin.  We
therefore exclude BNS and make no CBC-complete or astrophysical-rate claim.

\section{Discussion}
\label{sec:discussion}

The coherent DSD removes one avoidable ambiguity---a mismatch between the
representation used to build the index and the one used to query it---but it
does not make the survivor label physically self-interpreting.  Three distinct
uncertainties remain.

First, both the motivation and the effect of native adaptation are
detector-dependent.  The matched experiment resolves an O3b--O4a shift and a
native-score reduction for H1, whereas the corresponding L1 intervals include
zero.  Published O4a detector characterization provides plausible but
non-causal context: LHO transient noise was dominated by low-SNR 10--50\,Hz
features with a broadband source mitigated during O4a, whereas LLO transient
noise was dominated by seasonally varying, microseism-driven scattered light
\cite{soni2025}.  Our embedding-level experiment does not localize the
contributing bands or morphologies and therefore cannot attribute the measured
asymmetry to those mechanisms.  Where adaptation is used, it trades shift reduction against absorption:
a background index can regard a recurrent new morphology as normal precisely
because it is recurrent.  Candidate-vetoed sampling limits direct circularity,
but cannot remove unflagged members of that population.  The replicated
three-morphology experiment makes this a measurable, configuration-dependent
failure mode.

Second, preprocessing and dictionary construction are part of the statistical
model.  The moderate draw stability, $K$ dependence, and whitening-context
sensitivity show that representation choices must be included in uncertainty
analyses.  A deterministic forward pass is not equivalent to a robust
scientific disposition.

Third, physical nulls must match the nuisance structure.  In PEM coherence,
time shifts do not control persistent spectral coupling; a quiet zero-lag null
reduces 48 apparent positives to 10 confirmed endpoints.  In catalogue
matching, a raw overlap count is uninterpretable without processed coverage
and a time-structured null.  In cross-detector tests, embedding similarity is
not a substitute for light-travel-consistent strain correlation.

These findings also delimit the relationship to astrophysical anomaly
searches.  DANTE is useful as a detector-characterization triage system: it can
rank unusual windows, organize candidate inspection, and prioritize auxiliary
investigation.  The present calibration cannot supply astrophysical
significance.  A future search-oriented system would need injection populations
covering its target sources, causal or training-period-only adaptation,
multi-detector coherence in the ranking statistic, and explicit false-alarm
calibration over searched livetime.

\section{Reproducibility and limitations}

All class-dependent analyses load the representation-versioned taxonomy
`idxq4-64\_queryq4-64`, validate the index hash and Q-range, and reject missing
or non-finite scores. The complete repository suite contains 174 passing tests
and one platform-dependent skip.
A separate
fail-closed artefact audit checks ten named C2 stages and their upstream
calibration contracts.  The catalogue and PEM experiments have dedicated
input/output SHA256 manifests; new scans persist an append-only ledger of
successfully scored windows.

The main limitations are scientific rather than hidden execution failures.
The historical O4a scan lacks an exact successful-window ledger, contains ten
logged unreadable raw blocks, and physical-coincidence missingness is not known
to be random; PEM uses only publicly available, non-safety-certified channels;
the C2 study emphasizes deliberately near-boundary samples, while final
replicates use 160 candidates per population; the absorption matrix is
synthetic and restricted to O4a L1, one peak amplitude and one prescribed
duration per morphology, three
morphologies, and three clustering seeds; known-glitch controls are binary O3b
comparisons rather than O4a multiclass recall; the blind map has
eight trials per cell; and the CBC campaign excludes BNS.  The native index and
thresholds are O4a-specific, so another observing run requires new index
construction, calibration, and empirical validation.

\section{Conclusions}

A representation-coherent native calibration materially changes an
unsupervised O4a candidate taxonomy, but it does not convert statistical
novelty into a new glitch class.  The most defensible results are the measured
failure modes: moderate sensitivity to the native-background draw, dependence
on dictionary size and whitening context, a demonstrated absorption mechanism
for recurrent anomalies, and a morphology-dependent low-Q blind region.
Physical and
catalogue nulls resolve no \robust-class enrichment, cross-detector excess, or
catalogue overlap excess.  Compact-binary controls further show that run-native
adaptation can suppress signals flagged by a historical reference.

For unsupervised GW detector characterization, native recalibration can be
necessary when a shift is measured, but it is not neutral.  Credible discovery
requires publishing the
representation contract, calibration populations, negative controls, and
failure modes together with the candidate list.

\section*{Data availability}

Public O4a strain and auxiliary data are available from GWOSC
\cite{gwosc_aux_o4,gwosc_o4a_release}.  DANTE 3.7.0 is archived as a versioned
software snapshot on Zenodo \cite{dante_zenodo}.  The representation-versioned
result tables, per-trial outputs, environment records, SHA256 manifests, and
verification tools supporting this revision are archived separately in the
DANTE v6 reproducibility and validation evidence dataset
\cite{dante_v6_evidence}.  Their repository paths and provenance are enumerated
in the accompanying lab notebook.

\section*{Code availability}

The analysis code, command reference, tests, and paper-figure generator are
available in the DANTE repository associated with \cite{dante_zenodo}.  A new
run requires run-specific native indices, thresholds, and null calibrations.

\section*{Acknowledgements}

OpenAI Codex (GPT-5, service version accessed July--August 2026) was used for
language editing, consistency checks
between manuscript claims and saved artefacts, and assistance with test and
figure code.  It was not used to generate scientific data.  The author
designed the study, executed the analyses, checked the artefacts and
references, and accepts responsibility for the manuscript.

\section*{Funding}

This research received no external funding.

\section*{Competing interests}

The author declares no competing interests.

\section*{Author contributions}

L.C. conceived the study, developed the software, performed the analyses,
interpreted the results, and wrote and reviewed the manuscript.

\appendix

\section{Correction and claim-control log}

The present analysis corrects four interpretive issues.  (i) Catalogue timestamps
produced before 24 July 2026 label a padded fetch and precede the scored
32\,s window by 4\,s; scores are unchanged.  (ii) The physical re-run corrects
an axis mapping and yields 13/8,806 values above $\tau_{\rm cc}$.  A single
on-source value and per-event shifted maximum are not exchangeable, so this is
a conservative screen rather than a tail-count test.  (iii) DSD now uses a
coherent Q64/Q64 index/query contract; legacy Q32/Q64 classes are not used for
current claims.  (iv) Blind-spot injections are centred, and whitening sensitivity
must reproduce the pad-4 stored-score anchor.

Withdrawn claims include a universal absorption threshold, invariant ranking,
\robust-class PEM enrichment, survey-wide whitening stability, a universal
high-Q boundary, CBC-complete efficiency, catalogue recall, and preceding rate limits.
The latter no longer share a valid funnel after 4,676 paired dispositions
change.  The final PEM two-null endpoint reduces 48 time-shift exceedances to
10 confirmed endpoints after the quiet zero-lag control.

\clearpage
\bibliographystyle{apsrev4-2}
\bibliography{references}

@article{LIGOScientific:2014pky,
  author  = {Aasi, J. and others},
  title   = {{Advanced LIGO}},
  journal = {Class. Quantum Grav.},
  volume  = {32},
  pages   = {074001},
  year    = {2015},
  doi     = {10.1088/0264-9381/32/7/074001}
}

@article{Acernese_2014,
  author  = {Acernese, F. and others},
  title   = {{Advanced Virgo: a second-generation interferometric gravitational wave detector}},
  journal = {Class. Quantum Grav.},
  volume  = {32},
  pages   = {024001},
  year    = {2015},
  doi     = {10.1088/0264-9381/32/2/024001}
}

@article{KAGRA:2020tym,
  author  = {Akutsu, T. and others},
  title   = {{Overview of KAGRA: Detector design and construction history}},
  journal = {Prog. Theor. Exp. Phys.},
  volume  = {2021},
  pages   = {05A101},
  year    = {2021},
  doi     = {10.1093/ptep/ptaa125}
}

@article{robinet2020,
  author  = {Robinet, F. and Arnaud, N. and Leroy, N. and Lundgren, A. and Macleod, D. and McIver, J.},
  title   = {{Omicron: a tool to characterize transient noise in gravitational-wave detectors}},
  journal = {SoftwareX},
  volume  = {12},
  pages   = {100620},
  year    = {2020},
  doi     = {10.1016/j.softx.2020.100620}
}

@article{smith2011,
  author  = {Smith, J. R. and Abbott, T. and Hirose, E. and Leroy, N. and Macleod, D. and McIver, J. and Saulson, P. and Shawhan, P.},
  title   = {{A hierarchical method for vetoing noise transients in gravitational-wave detectors}},
  journal = {Class. Quantum Grav.},
  volume  = {28},
  pages   = {235005},
  year    = {2011},
  doi     = {10.1088/0264-9381/28/23/235005}
}

@article{cornish2015,
  author  = {Cornish, N. J. and Littenberg, T. B.},
  title   = {{BayesWave: Bayesian inference for gravitational wave bursts and instrument glitches}},
  journal = {Class. Quantum Grav.},
  volume  = {32},
  pages   = {135012},
  year    = {2015},
  doi     = {10.1088/0264-9381/32/13/135012}
}

@article{nuttall2018,
  author  = {Nuttall, L. K.},
  title   = {{Characterizing transient noise in the LIGO detectors}},
  journal = {Phil. Trans. R. Soc. A},
  volume  = {376},
  pages   = {20170286},
  year    = {2018},
  doi     = {10.1098/rsta.2017.0286}
}

@article{davis2021,
  author  = {Davis, D. and Areeda, J. S. and Berger, B. K. and others},
  title   = {{LIGO detector characterization in the second and third observing runs}},
  journal = {Class. Quantum Grav.},
  volume  = {38},
  pages   = {135014},
  year    = {2021},
  doi     = {10.1088/1361-6382/abfd85}
}

@article{soni2025,
  author  = {Soni, S. and Berry, C. P. L. and Coughlin, S. B. and others},
  title   = {{LIGO Detector Characterization in the first half of the fourth Observing run}},
  journal = {arXiv preprint},
  year    = {2025},
  eprint  = {2409.02831},
  archiveprefix = {arXiv}
}

@article{powell2015,
  author  = {Powell, J. and Trifir\`{o}, D. and Cuoco, E. and others},
  title   = {{Classification methods for noise transients in advanced gravitational-wave detectors}},
  journal = {Class. Quantum Grav.},
  volume  = {32},
  pages   = {215012},
  year    = {2015},
  doi     = {10.1088/0264-9381/32/21/215012}
}

@article{pankow2018,
  author  = {Pankow, C. and Chatziioannou, K. and Chase, E. A. and others},
  title   = {{Mitigation of the instrumental noise transient in gravitational-wave data surrounding GW170817}},
  journal = {Phys. Rev. D},
  volume  = {98},
  pages   = {084016},
  year    = {2018},
  doi     = {10.1103/PhysRevD.98.084016}
}

@article{zevin2017,
  author  = {Zevin, M. and Coughlin, S. and Bahaadini, S. and others},
  title   = {{Gravity Spy: integrating advanced LIGO detector characterization, machine learning, and citizen science}},
  journal = {Class. Quantum Grav.},
  volume  = {34},
  pages   = {064003},
  year    = {2017},
  doi     = {10.1088/1361-6382/aa5cea}
}

@article{glanzer2023,
  author  = {Glanzer, J. and Banagiri, S. and Coughlin, S. B. and others},
  title   = {{Data quality up to the third observing run of Advanced LIGO: Gravity Spy glitch classifications}},
  journal = {Class. Quantum Grav.},
  volume  = {40},
  pages   = {065004},
  year    = {2023},
  doi     = {10.1088/1361-6382/acb633}
}

@article{oquab2024,
  author  = {Oquab, M. and Darcet, T. and Moutakanni, T. and others},
  title   = {{DINOv2: Learning Robust Visual Features without Supervision}},
  journal = {Trans. Mach. Learn. Res.},
  year    = {2024},
  eprint  = {2304.07193},
  archiveprefix = {arXiv}
}

@inproceedings{darcet2024,
  author    = {Darcet, T. and Oquab, M. and Doup\'{e}, E. and Bourdoukan, R.},
  title     = {{Vision Transformers Need Registers}},
  booktitle = {Proc. ICLR},
  year      = {2024},
  eprint    = {2309.16588},
  archiveprefix = {arXiv}
}

@misc{gwosc_aux_o4,
  author = {{Gravitational Wave Open Science Center}},
  title  = {{O4 Auxiliary Channel Data Release}},
  year   = {2025},
  doi    = {10.7935/kt51-6n86},
  url    = {https://gwosc.org/auxiliary/}
}

@article{gwosc2023,
  author  = {Abbott, R. and others},
  title   = {{Open Data from the Third Observing Run of LIGO, Virgo, KAGRA, and GEO}},
  journal = {Astrophys. J. Suppl. Ser.},
  volume  = {267},
  pages   = {29},
  year    = {2023},
  doi     = {10.3847/1538-4365/acdc9f}
}

@misc{dante_zenodo,
  author = {Cirfeta, L.},
  title  = {{DANTE (Domain-Adaptive Network for Transient Evaluation)}},
  year   = {2026},
  version = {3.7.0},
  doi    = {10.5281/zenodo.21912589},
  howpublished = {\url{https://doi.org/10.5281/zenodo.21912589}},
  note   = {Version 3.7.0},
  publisher = {Zenodo}
}

@misc{dante_prior,
  author = {Cirfeta, L.},
  title  = {{An Unsupervised Search for Novel Instrumental Glitches in LIGO O4a: Multi-Scale Sensitization, Empirical Physical Vetoes, and Rate Upper Limits}},
  howpublished = {arXiv:2607.18136},
  year   = {2026},
  eprint = {2607.18136},
  archiveprefix = {arXiv}
}

@misc{dante_v6_evidence,
  author = {Cirfeta, L.},
  title  = {{DANTE v6: Reproducibility and Validation Evidence Bundle}},
  year   = {2026},
  version = {6},
  doi    = {10.5281/zenodo.21925453},
  howpublished = {\url{https://doi.org/10.5281/zenodo.21925453}},
  note   = {Version 6 dataset},
  publisher = {Zenodo}
}

@article{welch1967,
  author  = {Welch, P. D.},
  title   = {{The use of fast Fourier transform for the estimation of power spectra}},
  journal = {IEEE Trans. Audio Electroacoust.},
  volume  = {15},
  pages   = {70--73},
  year    = {1967},
  doi     = {10.1109/TAU.1967.1161901}
}

@article{chatterji2004,
  author  = {Chatterji, S. and Blackburn, L. and Martin, G. and Katsavounidis, E.},
  title   = {{Multiresolution techniques for the detection of gravitational-wave bursts}},
  journal = {Class. Quantum Grav.},
  volume  = {21},
  pages   = {S1809--S1818},
  year    = {2004},
  doi     = {10.1088/0264-9381/21/20/024},
  eprint  = {gr-qc/0412119},
  archiveprefix = {arXiv}
}

@article{essick2020,
  author  = {Essick, R. and Godwin, P. and Hanna, C. and Blackburn, L. and Katsavounidis, E.},
  title   = {{iDQ: Statistical inference of non-Gaussian noise with auxiliary degrees of freedom in gravitational-wave detectors}},
  journal = {Mach. Learn.: Sci. Technol.},
  volume  = {2},
  pages   = {015004},
  year    = {2020},
  doi     = {10.1088/2632-2153/abab5f}
}

@article{cuoco2021,
  author  = {Cuoco, E. and Powell, J. and Cavaglia, M. and others},
  title   = {{Enhancing Gravitational-Wave Science with Machine Learning}},
  journal = {Mach. Learn.: Sci. Technol.},
  volume  = {2},
  pages   = {011002},
  year    = {2021},
  doi     = {10.1088/2632-2153/abb93a}
}

@article{razzano2018,
  author  = {Razzano, M. and Cuoco, E.},
  title   = {{Image-based deep learning for classification of noise transients in gravitational wave detectors}},
  journal = {Class. Quantum Grav.},
  volume  = {35},
  pages   = {095016},
  year    = {2018},
  doi     = {10.1088/1361-6382/aab793}
}

@article{coughlin2019,
  author  = {Coughlin, S. and others},
  title   = {{Classifying the unknown: discovering novel gravitational-wave detector glitches using similarity learning}},
  journal = {Phys. Rev. D},
  volume  = {99},
  pages   = {082002},
  year    = {2019},
  doi     = {10.1103/PhysRevD.99.082002}
}

@article{george2018,
  author  = {George, D. and Shen, H. and Huerta, E. A.},
  title   = {{Classification and unsupervised clustering of LIGO data with Deep Transfer Learning}},
  journal = {Phys. Rev. D},
  volume  = {97},
  pages   = {101501},
  year    = {2018},
  doi     = {10.1103/PhysRevD.97.101501},
  eprint  = {1711.07468},
  archiveprefix = {arXiv}
}

@misc{gspy_zenodo,
  author    = {Glanzer, J. and Coughlin, S. and Bahaadini, S. and others},
  title     = {{Gravity Spy: glitch classifications for Advanced LIGO O1--O3}},
  year      = {2021},
  doi       = {10.5281/zenodo.5649212},
  publisher = {Zenodo}
}

@misc{lvk_burst_o4a,
  author  = {{LIGO Scientific Collaboration, Virgo Collaboration and KAGRA Collaboration}},
  title   = {All-sky search for short gravitational-wave bursts in the first part of the fourth LIGO-Virgo-KAGRA observing run},
  year    = {2025},
  eprint  = {2507.12374},
  archiveprefix = {arXiv},
  primaryclass = {gr-qc}
}

@misc{gwosc_o4a_release,
  author  = {{LIGO Scientific Collaboration, Virgo Collaboration and KAGRA Collaboration}},
  title   = {Open Data from LIGO, Virgo, and KAGRA through the First Part of the Fourth Observing Run},
  year    = {2025},
  eprint  = {2508.18079},
  archiveprefix = {arXiv},
  primaryclass = {gr-qc}
}

@article{biswas2013,
  author  = {Biswas, R. and Blackburn, L. and Cao, J. and others},
  title   = {{Application of machine learning algorithms to the study of noise artifacts in gravitational-wave data}},
  journal = {Phys. Rev. D},
  volume  = {88},
  pages   = {062003},
  year    = {2013},
  doi     = {10.1103/PhysRevD.88.062003}
}

@article{mukund2017,
  author  = {Mukund, N. and Abraham, S. and Kandhasamy, S. and Mitra, S. and Philip, N. S.},
  title   = {{Transient classification in LIGO data using difference boosting neural network}},
  journal = {Phys. Rev. D},
  volume  = {95},
  pages   = {104059},
  year    = {2017},
  doi     = {10.1103/PhysRevD.95.104059}
}

@article{bahaadini2018,
  author  = {Bahaadini, S. and Noroozi, V. and Rohani, N. and others},
  title   = {{Machine learning for Gravity Spy: Glitch classification and dataset}},
  journal = {Inf. Sci.},
  volume  = {444},
  pages   = {172--186},
  year    = {2018},
  doi     = {10.1016/j.ins.2018.02.068}
}

@article{vajente2020,
  author  = {Vajente, G. and Huang, Y. and Isi, M. and Driggers, J. C. and Kissel, J. S. and Szczepa{\'n}czyk, M. J. and Vitale, S.},
  title   = {{Machine-learning nonstationary noise out of gravitational-wave detectors}},
  journal = {Phys. Rev. D},
  volume  = {101},
  pages   = {042003},
  year    = {2020},
  doi     = {10.1103/PhysRevD.101.042003}
}

@article{ormiston2020,
  author  = {Ormiston, R. and Nguyen, T. and Coughlin, M. and Adhikari, R. X. and Katsavounidis, E.},
  title   = {{Noise reduction in gravitational-wave data via deep learning}},
  journal = {Phys. Rev. Research},
  volume  = {2},
  pages   = {033066},
  year    = {2020},
  doi     = {10.1103/PhysRevResearch.2.033066}
}

@misc{sakai2022,
  author        = {Sakai, Y. and Itoh, Y. and Jung, P. and others},
  title         = {{Training Process of Unsupervised Learning Architecture for Gravity Spy Dataset}},
  year          = {2022},
  eprint        = {2208.03623},
  archiveprefix = {arXiv}
}

@misc{laguarta2023,
  author        = {Laguarta, P. and van der Laag, R. and Lopez, M. and others},
  title         = {{Detection of anomalies amongst LIGO's glitch populations with autoencoders}},
  year          = {2023},
  eprint        = {2310.03453},
  archiveprefix = {arXiv}
}

@misc{raikman2023,
  author        = {Raikman, R. and Moreno, E. A. and Govorkova, E. and others},
  title         = {{GWAK: Gravitational-Wave Anomalous Knowledge with Recurrent Autoencoders}},
  year          = {2023},
  eprint        = {2309.11537},
  archiveprefix = {arXiv}
}

@misc{wu2024,
  author        = {Wu, Y. and Zevin, M. and Berry, C. P. L. and others},
  title         = {{Advancing Glitch Classification in Gravity Spy: Multi-view Fusion with Attention-based Machine Learning for Advanced LIGO's Fourth Observing Run}},
  year          = {2024},
  eprint        = {2401.12913},
  archiveprefix = {arXiv}
}

\end{document}